\documentclass[aps,prl,twocolumn,superscriptaddress,amsmath,amssymb,graphicx]{revtex4}

\newif\ifstructure
\structuretrue

\providecommand{\topicsentence}[1]{{\color{black} #1 }}

\usepackage{appendix}
\usepackage{amsmath}

\newcommand{\beq}{\begin{equation}}
\newcommand{\eeq}{\end{equation}}
\newcommand{\bea}{\begin{eqnarray}}
\newcommand{\eea}{\end{eqnarray}}
\providecommand{\abs}[1]{\left\lvert#1\right\rvert}

\providecommand{\bra}[1]{\langle #1 \rvert}
\providecommand{\ket}[1]{\lvert #1 \rangle}

\newcommand{\ketbra}[2]{\left| {#1} \right\rangle\left\langle {#2}\right|}

\usepackage{afterpage}
\usepackage{empheq}

\providecommand{\bra}[1]{\langle #1 \rvert}
\providecommand{\ket}[1]{\lvert #1 \rangle}

\newcommand{\Leff}{L_{\text{mixed}}}
\newcommand{\Ueff}{U_{\text{mixed}}}

\newcommand{\Lnh}{L_{\text{NH}}}
\newcommand{\LD}{L_{\text{Lindblad}}}

\newcommand{\sP}{\mathcal{P}}
\newcommand{\sQ}{\mathcal{Q}}

\usepackage{subcaption}
\usepackage{tikz}
\usepgflibrary{arrows}
\usetikzlibrary{decorations}
\usetikzlibrary{decorations.pathmorphing,patterns}
\usetikzlibrary{scopes}
\usetikzlibrary{shapes, matrix}

\providecommand{\boxG}{\fcolorbox{black}{green}{\rule{0pt}{4pt}\rule{4pt}{0pt}}}
\providecommand{\boxS}{\fcolorbox{black}{red}{\rule{0pt}{4pt}\rule{4pt}{0pt}}}
\providecommand{\boxB}{\fcolorbox{black}{blue}{\rule{0pt}{4pt}\rule{4pt}{0pt}}}

\usepackage{tikz}
\usepgflibrary{arrows}
\usetikzlibrary{decorations}
\usetikzlibrary{decorations.pathmorphing,patterns}
\usetikzlibrary{scopes}
\usetikzlibrary{shapes, matrix}
\usepackage{xcolor}

\begin{document}

\title{A continuous transformation between non-Hermitian Hamiltonian and Lindbladian evolution}

\author{Daniel Finkelstein-Shapiro}
\email{daniel.finkelstein@iquimica.unam.mx}
\affiliation{Instituto de Qu\'{i}mica, Universidad Nacional Aut\'{o}noma de M\'{e}xico, CDMX, M\'{e}xico}

\begin{abstract}
Non-Hermitian Hamiltonians and Lindblad operators are some of the most important generators of dynamics for describing quantum systems interacting with different kinds of environments. 
The first type differs from conservative evolution by an anti-Hermitian term that causes particle decay, while the second type differs by a dissipation operator in Lindblad form that allows energy exchange with a bath. 
However, although under some conditions the two types of maps can be used to describe the same observable, they form a disjoint set.   
In this work, we propose a generalized generator of dynamics of the form $\Leff(z,\rho_S) = -i[H,\rho_S] + \sum_i \left(\frac{\Gamma_{c,i}}{z+\Gamma_{c,i}}F_i\rho_S F_i^{\dagger} -\frac{1}{2} \{F_i^{\dagger} F_i,\rho_S \}_+\right)$ that depends on a general energy $z$, and has a tunable parameter $\Gamma_c$ that determines the degree of particle density lost. It has as its limits non-Hermitian ($\Gamma_c \to 0$) and Lindbladian dynamics ($\Gamma_c \to \infty$). The intermediate regime evolves density matrices such that $0 \leq \text{Tr} (\rho_S) \leq 1$. We derive our generator with the help of an ancillary continuum manifold acting as a sink for particle density. The evolution describes a system that can exchange both particle density and energy with its environment. We illustrate its  features for a two level system and a five $M$ level system with a coherent population trapping point. 
\end{abstract}

\maketitle


\section{Introduction}

\topicsentence{Master equations for open quantum systems allow an efficient explicit description of a system interacting with an environment we need not describe explicitly. Two very important classes of master equations correspond to 1) particle density exchange with an environment and 2) energy exchange with a bath.}   

A system exchanges particle density with its environment in a molecular junction where the electron goes from molecule (the system) to lead (the environment), when the electron of an atom or molecule (the system) photoionizes into states of the continuum (the environment), or in a waveguide (the system) which is pumped, and also leaks photons (from and to an environment). 
The extended Hilbert space for such a system $\mathcal{H}_S$ and environment $\mathcal{H}_E$ is $\mathcal{H}=\mathcal{H}_S \oplus \mathcal{H}_E$, the projection onto the system space is done via Feshbach projectors \cite{Feshbach1962}, and the resulting master equation for the system is a Hamiltonian evolution with a non-Hermitian (NH) term. The resulting description captures a wealth of non-trivial phenomena notably exceptional points  \cite{Moiseyev2011,Rotter2005}.
NH Hamiltonians have found important uses in spectroscopy \cite{Mukamel1995,Roccati2022} and  plasmonic systems \cite{FinkelsteinShapiro2021,Delga2014}. 

When the subsystem is exchanging energy with a bath, the Hilbert space of the entire system is $\mathcal{H}=\mathcal{H}_S \otimes \mathcal{H}_B$ Contrary to the NH case, here, the number of particles in $\mathcal{H}_S$ is conserved. 
The state of the system is described by the density matrix $\rho_S$, and in the case of a Markovian bath, the operator describing its evolution is of Lindblad-GKS form \cite{Lindblad1976, Gorini1976}. This equation has been extensively used for studying dissipative dynamics in condensed phase, photosynthetic systems, quantum information, and dissipative phase transitions \cite{May2011,Valkunas2013}. 

The NH Hamiltonian evolution can be written using the density matrix formalism (setting $\hbar = 1$) as $\dot{\rho_S} = \Lnh (\rho) = - i ( H_{\text{NH}}\rho_S -\rho_S H_{\text{NH}}^\dagger)$ where $H_{\text{NH}} = H-\frac{i}{2}\sum_i F_i^{\dagger}F_i$, $H = H^{\dagger}$ and $F_i$ are the jump operators. The Lindblad equation is $\dot{\rho_S} = \Lnh (\rho_S) + \sum_i F_i \rho_S F_i^{\dagger}$. The term $F_i \rho_S F_i^{\dagger}$ restores the population destroyed by the non-Hermitian term back into the system so that for Lindbladian evolution $\text{Tr}(\rho_S(t))=1$ for all times. When it is missing - for NH Hamiltonian evolution - and with the exception of systems where loss and gain are balanced, the trace of $\rho_S$ diminishes in time so that in the limit $t \to \infty$ the trace vanishes.  
Thus, there is a discontinuous transition for the value of the long time limit of the trace of the density matrix, from zero to one. When $F_i\rho_S  F_i^{\dagger}$ is included the trace is preserved, when it is absent the trace decays to zero. 

There have been several works striving to compare the evolution with and without this trace restoring term. A non-Hermitian Hamiltonian was modified to create a trace-preserving map by compensating the loss of trace $\frac{d}{dt}(\text{Tr}(\rho_S(t)))=\frac{2}{i \hbar} \text{Tr}(\rho_S(t)F_i^{\dagger}F_i)$ \cite{Dorje2012,EcheverriArteaga2018,EcheverriArteaga2018a,EcheverriArteaga2019,Zloshchastiev2014}. The extra term added is nonlinear and brings forth interesting physics such as anharmoniticities in two-level systems \cite{Zloshchastiev2014}. 
Recently, Minganti et al. analyzed and compared the exceptional points in non-Hermitian Hamiltonians and Lindbladians, and investigated a connection between exceptional points by gradually turning on the term $F_i\rho F_i^{\dagger}$ by weighting it with a pre-factor $0<q<1$ \cite{Minganti2019,Minganti2020}. The authors separated exceptional points into purely non-Hermitian Hamiltonian, purely Lindbladian and common to both. However, the trace for this type of connection is also either 1 ($q=1$) or 0 ($q<1$). This modification results in a generator for non-conservative dynamical semigroups, which has also been used in describing heavy-ion dissipative collisions \cite{Alicki2007,Alicki1982}. 
 
We present in this work a connection between both NH and Lindblad evolutions such that the trace can vary continuously from zero to one. As such, it includes and goes beyond the previously studied non-conservative dynamical semigroups. The microscopic model and type of dynamics is different from earlier works and is based on dissipative systems with Hamiltonians with discrete and continuous  spectra. 
After stating the main result, we present the derivation of the map by means of an ancillary continuum. We then illustrate its main features with two examples, a two level system and an $M$-level system where coherent population trapping is possible. The connection is entirely general and valid whenever a Lindblad operator can be written.

\section{Main result}

We construct an evolution operator $U_{\text{mixed}}(t)$ that can connect the dynamics generated by non-Hermitian Hamiltonians
and that generated by Lindblad operators
This map is (setting $\hbar = 1$):
\begin{equation}
    U_{\text{mixed}}(t) = \frac{1}{2\pi i}\oint  \frac{e^{zt}}{z-L_{\text{mixed}}(z)} dz
    \label{eq:U_mixed}
\end{equation}
where
\begin{equation}
\begin{split}
    L_\text{mixed}(z,\rho_S) &= -i[H,\rho_S]\\
    &+\sum_i \left[ \Delta_i(z) F_i\rho_S F_i^{\dagger} -\frac{1}{2} \{F_i^{\dagger} F_i,\rho_S \}_+ \right]
    \end{split}
    \label{eq:main_result}
\end{equation}
where $H$ is a (Hermitian) Hamiltonian, $F_i$ are jump operators in the Lindblad dissipator, $\{\}_+$ is the anti-commutator and we have introduced a new $z-$dependent function  $\Delta_i(z) = \frac{\Gamma_{c,i}}{z+\Gamma_{c,i}}$. $\Gamma_{c,i}$ are new parameters that tune the fraction of Lindbladianity and non-Hermiticity.
The operator $L_\text{mixed}$ is $z-$dependent, which allows it to capture the loss of the trace of the density matrix in continuous amounts between 0 and 1. 
The factor $\Delta(z)$ becomes $1$ for Lindblad generators ($\Gamma_c \to \infty$) and $0$ for non-Hermitian Hamiltonian generators ($\Gamma_c \to 0$). For the limits to be true, $\Gamma_c$ has to be much bigger (smaller) than the largest (smallest) magnitude of the real part of the eigenvalues of the Lindblad (Non-Hermitian) operator. 
We will show that the evolution operator can be written compactly as
\begin{equation}
    U_\text{mixed}(t) = \sum_{i=1}^K \mathcal{X}_ie^{\lambda_i t} 
    \label{eq:second_main_result}
\end{equation}
where $\lambda_i$ and $\mathcal{X}_i$ are eigenvalues and projection operators that solve a nonlinear eigenvalue problem. Eq. \eqref{eq:second_main_result} differs from the exponential map of a standard generator in that, here, for an $N$ level system $K>N^2$, unlike for a Lindblad or NH operator where $K = N^2$. 
Eq. \eqref{eq:main_result} becomes a standard generator whenever it becomes z-independent. In the general case, it is a kernel akin to the memory kernels used in non-Markovian dynamics, expressed in the frequency domain and allowing exchange of particle density. 
Below, we outline the derivation of Eq.
(\ref{eq:U_mixed}-\ref{eq:second_main_result}). \newline

\section{Derivation}
\begin{widetext}
    
\begin{figure}
    \centering
    \includegraphics[width=1.0\textwidth]{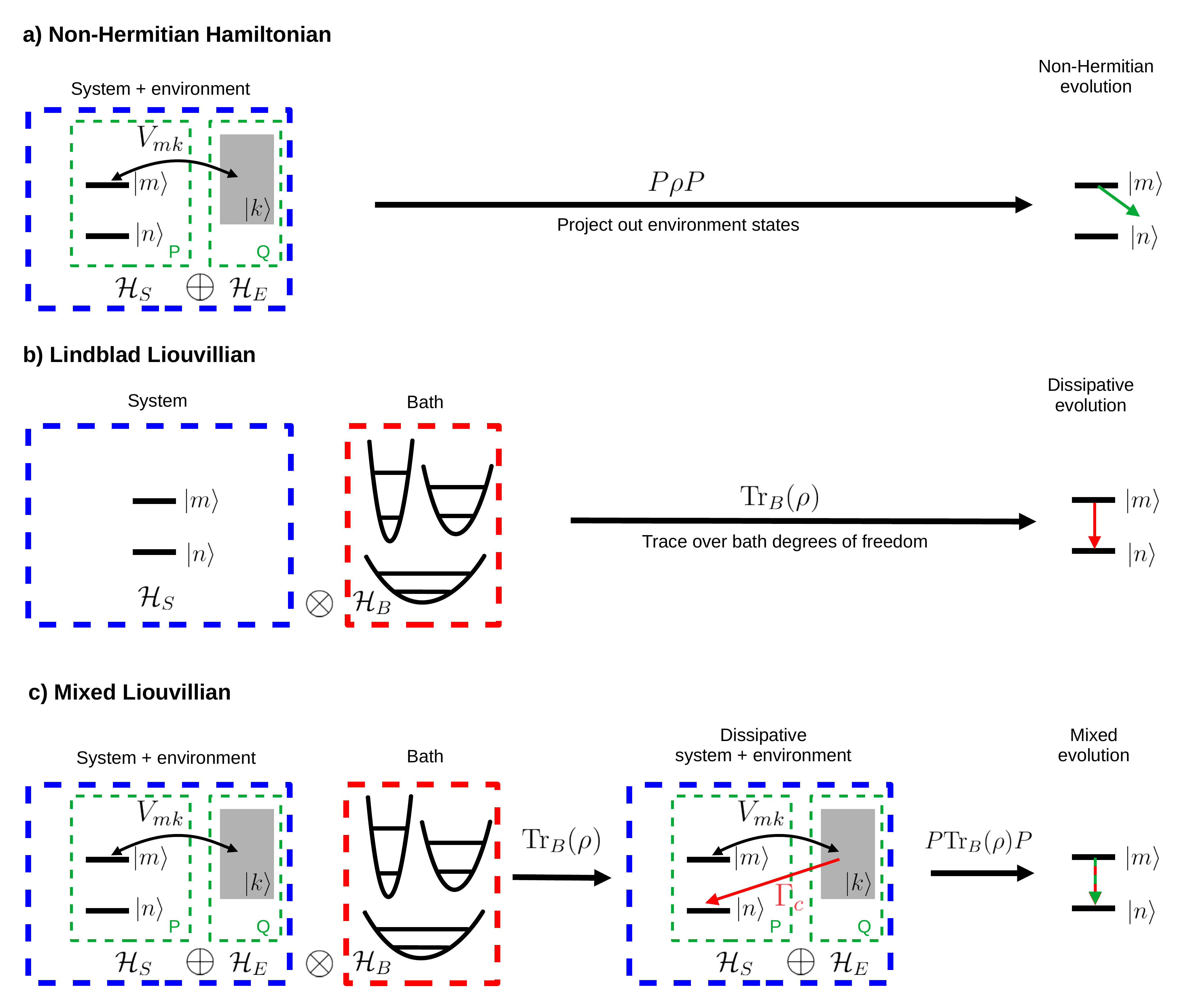}
    \caption{Microscopic models for different types of evolution. a) A NH Hamiltonian is obtained by projecting out a continuum of states $\ket{k}$ (partition $Q$) that couples to discrete level $\ket{m}$ (partition $P$). The effect is to induce a decay of state $\ket{m}$. b) A Lindblad Liouvillian is obtained by tracing out a bath of harmonic oscillators under appropriate assumptions (see text). The bath can open several transitions and we focus here on a relaxation from $\ket{m}$ to $\ket{n}$. c) A mixed Liouvillian with an interpolation between the dynamics is obtained by coupling a Hamiltonian structure identical to a) to a bath that creates a relaxation from $\ket{k}$ to $\ket{n}$. Projecting out states $\ket{k}$ yields the mixed evolution. Double-sided arrows are Hamiltonian couplings, while red single-sided arrows are dissipative transitions.}
    \label{fig:microscopic_models}
\end{figure}

\end{widetext}

\textit{Microscopic models.} The maps considered in this work are reduced descriptions of a larger system, and have been obtained by tracing out or projecting out certain degrees of freedom \cite{Breuer2000}. These extended systems (or microscopic models) that produce NH Hamiltonians, Lindblad operators or a possible generalization are shown in Figure \ref{fig:microscopic_models}. To distinguish the models, we refer to additional degrees of freedom within a subsystem as the environment, and refer to other subsystems as the bath, although bath and environment are usually used interchangeably. 

One of the simplest ways to induce non-Hermiticity in a Hamiltonian is to connect discrete energy levels to an ancillary continuous manifold of states that acts as a particle sink and then remove this continuous manifold from the explicit description with Feshbach projectors (\cite{Feshbach1962,May2011,Finkelstein2018}, and Figure \ref{fig:microscopic_models}.a). For a continuous manifold obeying the wideband approximation \cite{Finkelstein2018}, this coupling to a continuum complexifies the energy of the coupled discrete state, i.e. for a single level $\ket{m}$ connected to a continuum of states through a coupling $V=\sqrt{\gamma/(2 \pi)}$, we get the transformation $H = \omega_m \ketbra{m}{m} \to H_{\text{NH}} = (\omega_m-i\gamma/2)\ketbra{m}{m}$. 
The Hilbert space of the system and ancillary manifold is $\mathcal{H}_S$ and $\mathcal{H}_E$, respectively, for a full Hilbert space $\mathcal{H}_S \oplus \mathcal{H}_E$. 
%
%
We define projectors $P$ and $Q$ which project onto the system or ancillary states, respectively, and calculate the system density matrix as $P \rho P = \rho_S$. The result is that $\dot{\rho}_S = -i(H_{\text{NH}} \rho_S - \rho_S H_{\text{NH}}^{\dagger})$ 
\cite{Feshbach1962,May2011,Finkelstein2018}.

The Lindblad form for dissipative evolution can be derived microscopically by coupling a system to a collection of harmonic oscillators in the Markovian limit. The respective Hilbert spaces of system and bath are $\mathcal{H}_S$ and $\mathcal{H}_B$ for a total Hilbert space $\mathcal{H}_S \otimes \mathcal{H}_B$. After tracing out the degrees of freedom of the bath under suitable approximations we obtain the additional term in the Liouville equation $\sum_i \left[F_i\rho_S F_i^{\dagger} -\frac{1}{2} \{F_i^{\dagger} F_i,\rho_S \}_+ \right]$. Each $F_i$ is a quantum jump operator for a dissipative channel with rate $\gamma_i$ (we include the rate in the jump operator, for example, for a transition from state $a$ to $b$ with rate $\gamma_{ab}$ we have $F=\sqrt{\gamma_{ab}}\ketbra{b}{a}$.). 
The lengthy derivation has been shown elsewhere and we do not reproduce it here (\cite{Breuer2000,Lindblad1976,Manzano2020}, Figure \ref{fig:microscopic_models}.c)).

We now construct a microscopic model that can reproduce the two previous dynamics using an ancillary continuum (Figure \ref{fig:microscopic_models}.c). 
We keep the idea of the NH model of destroying particle density by sending it to an ancillary continuous manifold, but then restore it from this manifold to a different discrete state via a dissipative pathway. 
The Hilbert spaces of the system, ancillary continuum and bath are $\mathcal{H}_S$, $\mathcal{H}_E$ and $\mathcal{H}_B$ respectively, so that the entire Hilbert space is $\mathcal{H}= (\mathcal{H}_S \oplus \mathcal{H}_E) \otimes \mathcal{H}_B $. The dynamics in $\mathcal{H}_S$ is obtained first by tracing out the degrees of freedom of the bath $\mathcal{H}_B$, and then projecting out the ancillary states $\mathcal{H}_E$. 
\newline



\textit{Solving for the evolution operator}. The Hamiltonian in $\mathcal{H}_S \otimes \mathcal{H}_E$ corresponding to Figure \ref{fig:microscopic_models}.c is:

\begin{equation}
\begin{split}
    H &= H_{\text{discrete}}+H_{\text{continuum}}\\
    &+H_{\text{discrete-continuum}} \\
    H_\text{discrete} &=  \omega_n\ketbra{n}{n}+\omega_m\ketbra{m}{m} \\
    H_\text{continuum} &= \int dk \omega_k \ketbra{k}{k} \\
    H_\text{discrete-continuum} &= \int dk (V_{mk}\ketbra{m}{k} + h.c.) \\
\end{split}
\end{equation}
We switch to Liouville space $\mathcal{H} \otimes \mathcal{H}^*$ (for a given Hilbert space $\mathcal{H}$) by means of the isomorphism $S_m \rho_S S_n^{\dagger} \to S_n^* \otimes S_m \rho_v$ where $\rho_v$ is a column-stretched vector built from the density matrix $\rho_S$ (we forego the subscript $v$ in the following and continue to use $\rho_S$ as the difference is clear from the context. We similarly use $\Leff$ for the operator in both Hilbert and Liouville spaces and rely on the context to distinguish them).  We can write the conservative part of the Liouvillian as $L_H = -i[1\otimes H - H^* \otimes 1]$. We add the dissipative transitions, resulting from tracing out the bath degrees of freedom, from the continuum manifold $\ket{k}$ to $n$ using the jump operator $F_{mn}^{(k)} = \sqrt{\Gamma_c} \ketbra{n}{k}$.
The Lindblad operator for discrete and continuous manifold is
\begin{equation}
L=L_H+ \int dk L_D(F_{mn}^{(k)})
\label{eq:full_L}
\end{equation}
with $L_D(F) = F^* \otimes F-\frac{1}{2} \left[ (F^{\dagger} F)^T \otimes 1 + 1 \otimes F^{\dagger} F \right] $. 

We next project out the ancillary continuous manifold of states. We have previously solved a similar system under steady-state conditions  \cite{Finkelstein2018,Finkelstein2016-1,Finkelstein-Shapiro2020}). 
The explicit form of the Feshbach projectors in Hilbert space are $P=\ketbra{n}{n}+\ketbra{m}{m}$ and $Q=\int dk \ketbra{k}{k}$, with $P+Q=1$. In Liouville space we write $\sP = P \otimes P$ and $\sQ = 1 - \sP$. To find the dynamics in $\sP$, we can calculate the evolution operator $\rho_S(t) = \sP \rho(t) = \sP U(t) \sP \rho(0) + \sP U(t) \sQ \rho(0)$ (remembering that $\rho$ is now a column-stretched vector in ($\mathcal{H}_S \oplus \mathcal{H}_E) \otimes (\mathcal{H}_S \oplus \mathcal{H}_E)^*$). As long as $\sQ \rho(0) = 0$, we only need 
\begin{equation}
    \sP U(t) \sP = \frac{1}{2\pi i} \oint dz \sP G(z) \sP e^{zt}  
    \label{eq:inverse_Laplace}
\end{equation}
where $\sP G(z) \sP = (z\sP-L_{\text{mixed}}(z))^{-1}$, and
\begin{equation}
\Leff(z) = \sP L \sP+\sP L \sQ G_0(z) \sQ L \sP
\end{equation}
where $L$ is the Lindblad operator in Equation \eqref{eq:full_L} and $\sQ G_0 \sQ = (z \sQ - \sQ L \sQ )^{-1}$. As we show in detail in the Appendix A, 
\begin{equation}
\begin{split}
    \Leff(z) &= \LD-\sum_i \frac{z}{z+\Gamma_{c,i}} J_i \\
    &= \Lnh+\sum_i \frac{\Gamma_{c,i}}{z+\Gamma_{c,i}} J_i
    \end{split}
    \label{eq:Leff}
\end{equation}
where $J_i = F_i^* \otimes F_i$ is the  operator that restores the lost population of the excited state back into the ground state, $\Lnh = -i[1\otimes H - H^*\otimes 1] - \sum_i \frac{1}{2}\left( 1\otimes F_i^{\dagger}F_i + (F_i^{\dagger}F_i)^T \otimes 1 \right)$ and $\LD = \Lnh + \sum_i F_i^* \otimes F_i$. This is the result stated in Eq. \eqref{eq:main_result}. 

Because of the $z-$dependence of $\Leff$, we cannot express the evolution operator as the exponential of $\Leff$. The generalized eigenvalue problem $(z-\Leff(z))\ket{v} =0$ is now nonlinear and in the case where $\Gamma_{c,i} \equiv \Gamma_c$ for all dissipative transitions $i$, it is quadratic. Eq. \eqref{eq:inverse_Laplace} can be numerically integrated to obtain the evolution. Instead, we solve it exactly in an extended space where the eigenvalue problem becomes linear (without any approximations). We write the explicit $z-$dependence of $\Leff(z)$ in the denominator as a quadratic pencil $D(z)$ \cite{Tisseur2001}:
\begin{equation}
    U(t) = \frac{1}{2\pi i}\oint \frac{(z+\Gamma_{c})e^{zt}}{ D(z) } dz
\end{equation}
where $D(z) = z^2 + A_1z+A_0$, $A_1 = \Gamma_c-\LD+J$ and $A_0=-\LD\Gamma_c$, with $J = \sum_i J_i$. The numerator of this resolvent now has a factor $(z+\Gamma_c)$. We solve the quadratic eigenvalue problem by doubling the dimension of Liouville's space and defining auxiliary matrices $\tilde{E}$, $\tilde{F}$ and $\tilde{M}$ \cite{Tisseur2001}:
\begin{equation}
    \begin{bmatrix}
    D(z) & 0 \\
    0 & 1
    \end{bmatrix} = \tilde{E}(z)(\tilde{M}-z \tilde{B}) \tilde{F}(z)
\end{equation}
where 
\begin{equation}
    \begin{split}
        \tilde{M} &= \begin{bmatrix}
        0 & 1 \\
        -A_0 & -A_1
        \end{bmatrix}, 
        \tilde{B} = \begin{bmatrix}
        1 & 0 \\
        0 & 1
        \end{bmatrix}, \\
        \tilde{E} &= \begin{bmatrix}
        (-A_1+z) & -1 \\
        1 & 0
        \end{bmatrix},
        \tilde{F} = \begin{bmatrix}
        1 & 0 \\
        z & 1
        \end{bmatrix} \\
    \end{split}
    \label{eq:M_matrix}
\end{equation}
We have written $1$ and $0$ as the identity and null matrix, respectively, with dimensions $N^2 \times N^2$ where $N$ is the number of levels of the subsystem we wish to describe explicitly (the 
$\mathcal{P}$ partition). 
We denote all operators in this extended space by a tilde, while the operators in the original Liouville space have no tilde. 
If $\tilde{E}$ and $\tilde{F}$ have non-zero determinants, the eigenvalues of $D(z)$ coincide with those of $\tilde{M}$ \cite{Tisseur2001}. 
We can calculate the determinants of $\tilde{E}$ and $\tilde{F}$ using the identity for block matrices $\det \left(\begin{bmatrix}
    A & B \\ C & D
\end{bmatrix} \right) = \det(A) \det(D-CA^{-1}B) $ which is valid as long as $A$ is invertible. This immediately gives $\det(\tilde{F})=1$, and $\det(\tilde{E})=\det(-A_1+z)\det([-A_1+z]^{-1})=1$ as long as $-A_1+z$ is invertible. 
The sought after inverse $(z-D(z))^{-1}$ is
\begin{equation}
    (z-D(z))^{-1} = S_o^T (z-\tilde{M})^{-1} S_e
    \label{eq:inverse_equivalent}
    \end{equation}
where we define the projection operator onto the original Hilbert space $S_o = \begin{bmatrix}
1 \\ 0
\end{bmatrix}$ and onto the extended space $S_e = \begin{bmatrix}
0 \\ 1
\end{bmatrix}$. 
We decompose the resolvent $\tilde{\mathcal{G}}(z)=(z-\tilde{M})^{-1}$ into its projectors \cite{kato1995perturbation}
\begin{equation}
    \tilde{\mathcal{G}}(z) = \sum_{i=1}^{2N^2}\frac{z+\Gamma_c}{z-\lambda_i}\tilde{X}_i
    \label{eq:generalized_resolvent}
\end{equation}
where $\lambda_i$ are the eigenvalues of $\tilde{M}$ and $\tilde{X}_i=\ketbra{v_i}{w_i}$ the corresponding projection operators built from the right ($\tilde{M}\ket{v_i}=\lambda_i\ket{v_i}$) and left ($\bra{w_i}\tilde{M}=\bra{w_i}\lambda_i$) eigenvectors. 
Combining Eqs. \eqref{eq:inverse_equivalent} and \eqref{eq:generalized_resolvent}, we  arrive at a compact expression for the evolution operator (see Eq. \eqref{eq:second_main_result})
\begin{equation}
    \Ueff(t) = \sum_i^K (\lambda_i+\Gamma_c)e^{\lambda_i t} S_o^T {X}_i S_e = \sum_i^K \mathcal{X}_i e^{\lambda_i t}
\end{equation}
where we have defined the generalized projection operators $\mathcal{X}_i  =  (\lambda_i+\Gamma_c) S_o^T \tilde{X}_i S_e$. 
The power of using an extended space can now be appreciated. Instead of solving the nonlinear eigenvalue problem involving $D(z)$, we solve the eigenvalues of the matrix $M$. Some of this eigenvalues will be equal to $-\Gamma_c$, and in this case no real pole exists in the resolvent of Eq. \eqref{eq:generalized_resolvent}. It is not a problem to count them as poles since the corresponding generalized projection operators will vanish.  
The operator $\Ueff(t)$ has more dynamical variables than dimensions. Because of this, the projection operators in the reduced space $\mathcal{X}_i$ are not always orthogonal (they are orthogonal in the total extended space) and so $\Ueff(t)$ cannot be expressed in general as the exponential map of a $z-$independent generator of dynamics. This is the new mathematics required to obtain the generalized dynamical map. \newline

As we introduce an ancillary continuum for each dissipative pathway, we can in principle have a different value $\Gamma_{c,i}$ for each continuum that mediates the relaxation between any two levels (Eq. \eqref{eq:Leff}). 
For every distinct $\Gamma_{c,i}$, the order of the nonlinear eigenvalue problem increases by one. The problem is not quadratic anymore, and a thorough investigation on the role of choosing different values for the continuum decay will be discussed in subsequent work. We can also choose to have one continuum mediate more than one dissipative transitions instead of having one continuum per transition. In this case the ancillary continuum is responsible for generating coherent transitions as well. This commonly occurs in Fano interferences (see for example \cite{Finkelstein2018}) and can be handled by the derived expressions, although its physical meaning is still unclear in the context of the connection between non-Hermitian Hamiltonian and Lindblad maps.  \newline

The density matrix that is evolved by $U_\text{mixed}(t)$ corresponds to the projection of the density matrix evolving in the discrete level and continuous manifold of states onto the discrete states only. The extended density matrix in ($\sP$ + $\sQ$) evolves according to a Lindblad operator, and as such it is Hermitian $\rho=\rho^{\dagger} $, and $\text{Tr}(\rho)=1$. By design, then, the system density matrix $\rho_S = \mathcal{P} \rho$ has the properties of Hermiticity $\rho_S = \rho_S^{\dagger}$ and also that $\text{Tr}(\rho_S) \leq 1$ where the equality is only in the case where the entire density remains in $\mathcal{P}$ (i.e. in the Lindblad limit).



\section{Examples}

We illustrate the dynamics proposed with two examples. We first study a two level system as it allows us to illustrate in detail the properties of eigenvalues and projectors. We then approach a more complicated M-level system that has a coherent population trapping point. 

\subsection{Two-level system}

We use a two-level system Hamiltonian $H = \delta_e \ketbra{e}{e} + (V_{eg} \ketbra{e}{g}+ h.c.)$, where $\delta_e$ is the detuning, $V_{eg}$ the coupling between the two discrete states, and a dissipator in Lindblad form with $F_1 = \sqrt{\gamma} \ketbra{g}{e}$. We show the dynamics for different values of $\Gamma_c$ in two cases: off-resonance $\delta_e \gg V_{eg}$ and on-resonance $\delta_e \ll V_{eg}$ for the initial condition $\rho(0) = \rho_{ee}$ (Fig 2.a-l). 

Off-resonance (Fig. 2.a-c), the mixing of ground and excited states is minimal and we obtain a decay of the excited state onto the ground state. The trace of the system depends on the value of $\Gamma_c$. For $\Gamma_c \ll 1$ we are in the non-Hermitian limit and the system decays to zero, while in the $\Gamma_c \to \infty$ limit we are in the Lindblad limit and the trace is preserved. In between, the particle density can leave the system to later return, and reach a steady-state where $\text{Tr}(\rho(t\to\infty))<1$. We show in addition the normalized fidelity $\mathcal{F}_i  = \text{Tr}(\sqrt{ \sqrt{\rho_i}\rho  \sqrt{\rho_i}   })/\sqrt{\text{Tr}(\rho_i)\text{Tr}(\rho)}$ for $i=$Lindblad, NH to give a quantitative measure as to the character of the evolution. 

We expect in general that the number of poles of the resolvent $\tilde{\mathcal{G}}$ to be eight. However, the $(z+\Gamma_c)$ factor in the numerator can remove simple poles at $z=-\Gamma_c$. 
While the calculation of operators and its resulting dynamics does not require us to know the number of poles that are removed by the numerator, we can study it explicitly in the case of a two-level system. As we have shown in the Appendix B, for a two-level system with a single dissipative pathway the poles of the generalized resolvent is five, i.e. there are three eigenvalues at $-\Gamma_c$. It can be also shown that the algebraic multiplicity of $-\Gamma_c$ is $N^2$ minus the number of connected dissipative pathways. For a two-level system connected to a finite temperature bath with incoherent pumping and incoherent decay this corresponds to six distinct eigenvalues (Appendix B, Figure \ref{fig:evolution_finite_T_bath}) while for the case of pure dephasing we recover all 8 distinct eigenvalues (Appendix B, Figure \ref{fig:evolution_pure_dephasing}). 

We can appreciate the five poles for the two-level system in Figs. 2.d-f (black crosses) where we have also shown the Lindbladian (blue circles) and non-Hermitian Hamiltonian (red triangles) eigenvalues for comparison. 
%
The on-resonance case shows a qualitatively similar behavior regarding the trace, with a clear presence of Rabi oscillations and a near equal mixture of ground and excited state in the long-time limit. In contrast with the off-resonant case, the normalized fidelity $\mathcal{F}_\text{NH}$ does not represent the system at long times.   

Our description of non-Hermitian decay also involves a steady-state pole ($\lambda = 0$), which can be reconciled if the projector corresponding to this eigenvalue vanishes in the limit $\Gamma_c \to 0$. To investigate this in more detail, we 
study the structure of the eigenvalue manifold as a function of $\Gamma_c$ as well as the trace of each projector with an (arbitrary) initial state in the upper level $e$, $t_i = \text{Tr}(\mathcal{X}_i\rho_{ee})$. For clarity, we focus on $t_0(\lambda=0)$ and $\sum_i t_i(\lambda_i \neq 0)$ (see Figure 3 c,f,i). 
As we start from a normalized density matrix at time zero, we have $\sum_i t_i =1$, however depending on the value of $\Gamma_c$, this trace is carried by different projectors, and it is only $t_0$ that will survive in the long-time limit. As is expected, for dissipative evolution $\Gamma_c$, the trace is carried entirely by $t_0$, and for non-Hermitian decay it is carried by $t_5$, the fifth eigenvalue (Figs. 3.c,g,k).  
We also remark that the eigenvalue structure is not monotonic, but has avoided crossings and changes of symmetry as a function of the parameters (Figs. 3.a,d,g). The asymptotic value of the pole corresponding to the decay dynamics is not the same on resonance (goes to zero) or off resonance (reaches a finite value).

\begin{widetext}

\begin{figure}
    \centering
    \includegraphics[width=0.8\textwidth]{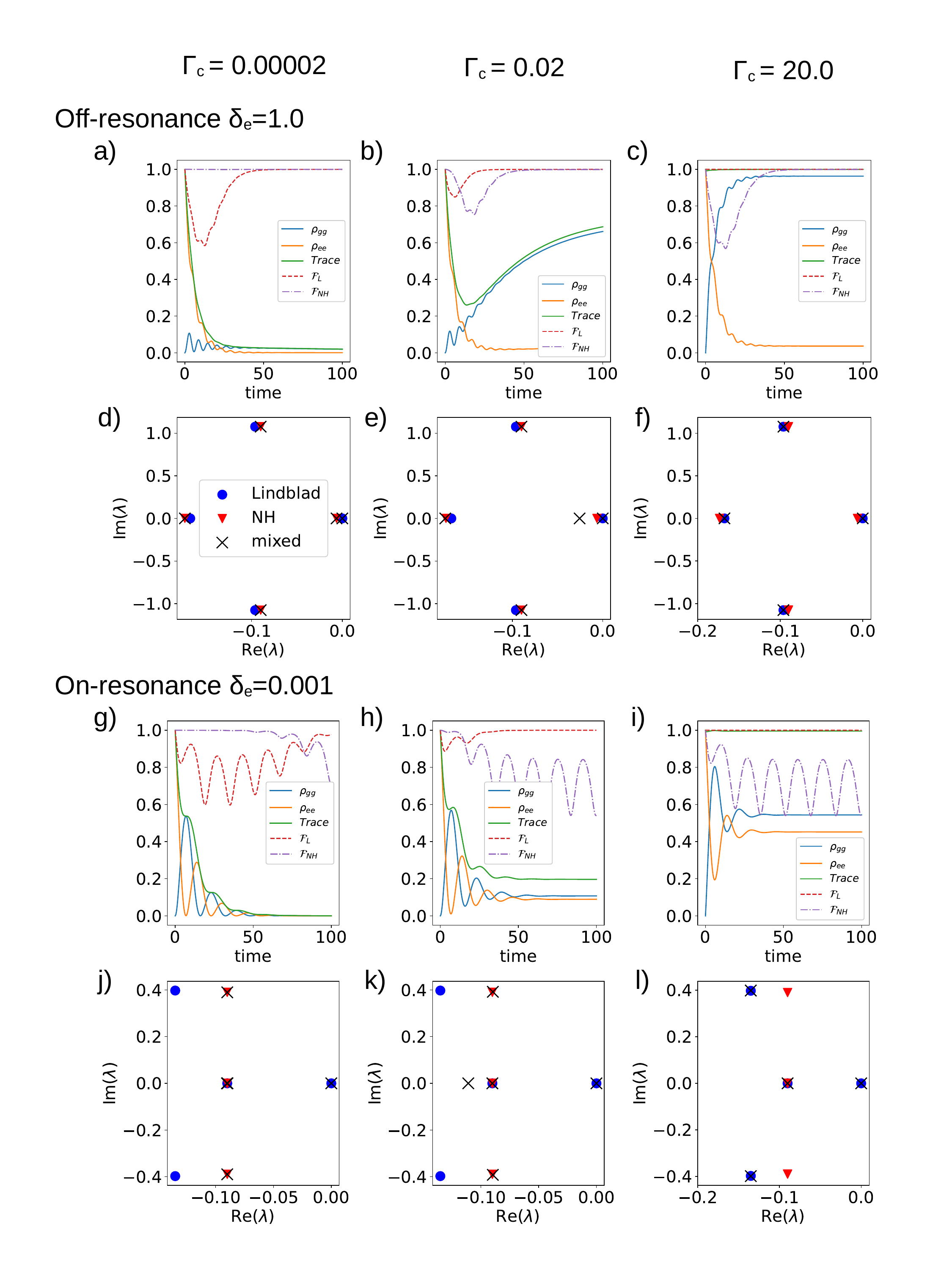}
    \caption{Analysis of a two level system off-resonance case $\delta_e=1.0$. The other parameters are $V_{eg}=0.2$, $\sqrt{\gamma/2\pi}=0.3$. Populations, trace and fidelity for values of $\Gamma_c=0.00002,0.02,20.0$ are shown in (a-c), respectively. The poles of the generalized resolvent of Eq. \eqref{eq:generalized_resolvent} are shown below ($x$), while the eigenvalues of the Lindblad and non-Hermitian operator are shown for comparison. g-l: same for the on-resonance case $\delta_e=0.001$. The fifth pole for $\Gamma_c=20.0$ is found at $\lambda \approx -\Gamma_c$ and is not shown.}
    \label{fig:dynamics}
\end{figure}

\end{widetext}

Previous work investigating the use of non-Hermitian Hamiltonians have highlighted the usefulness of the normalized density matrix $\rho_S(t)/\text{Tr}(\rho_S(t))$, whose element corresponding to the population of the excited state we plot in Fig 3. b,e,h. In consonance with the eigenvalue structure, we observe two different behaviours. For large $\Gamma_c$ the oscillations decay exponentially, while for small enough $\Gamma_c$ they are long-lived and do not decay exponentially. This corresponds to a regime where the coherences decay on the same timescale as the trace of the density matrix and so see their effective lifetime extended due to the normalization by the trace.

 Pure dephasing cannot be expressed in a purely non-Hermitian setup, however, the mixed map (Eq. \eqref{eq:main_result}) can be constructed for a pure dephasing operator ($F_1 = \sqrt{\gamma} \sigma_z$), and yields a physically valid evolution (Appendix B, Figure \ref{fig:evolution_pure_dephasing}). The non-Hermitian limit of pure dephasing operator corresponds to a term in the non-Hermitian operator $-\Gamma \rho_S$ where all elements of the density matrix decay with equal rate constant. On the other hand, $\mathcal{PT}$ symmetric non-Hermitian Hamiltonians pose problems for $\Gamma_c \neq 0$. This is because $\mathcal{PT}$ symmetry expressed with Lindblad operators requires the presence of negative dissipative rates which causes populations to become larger than one and smaller than zero. \newline

\begin{widetext}
    
\begin{figure}
    \centering
    \includegraphics[width=0.8\textwidth]{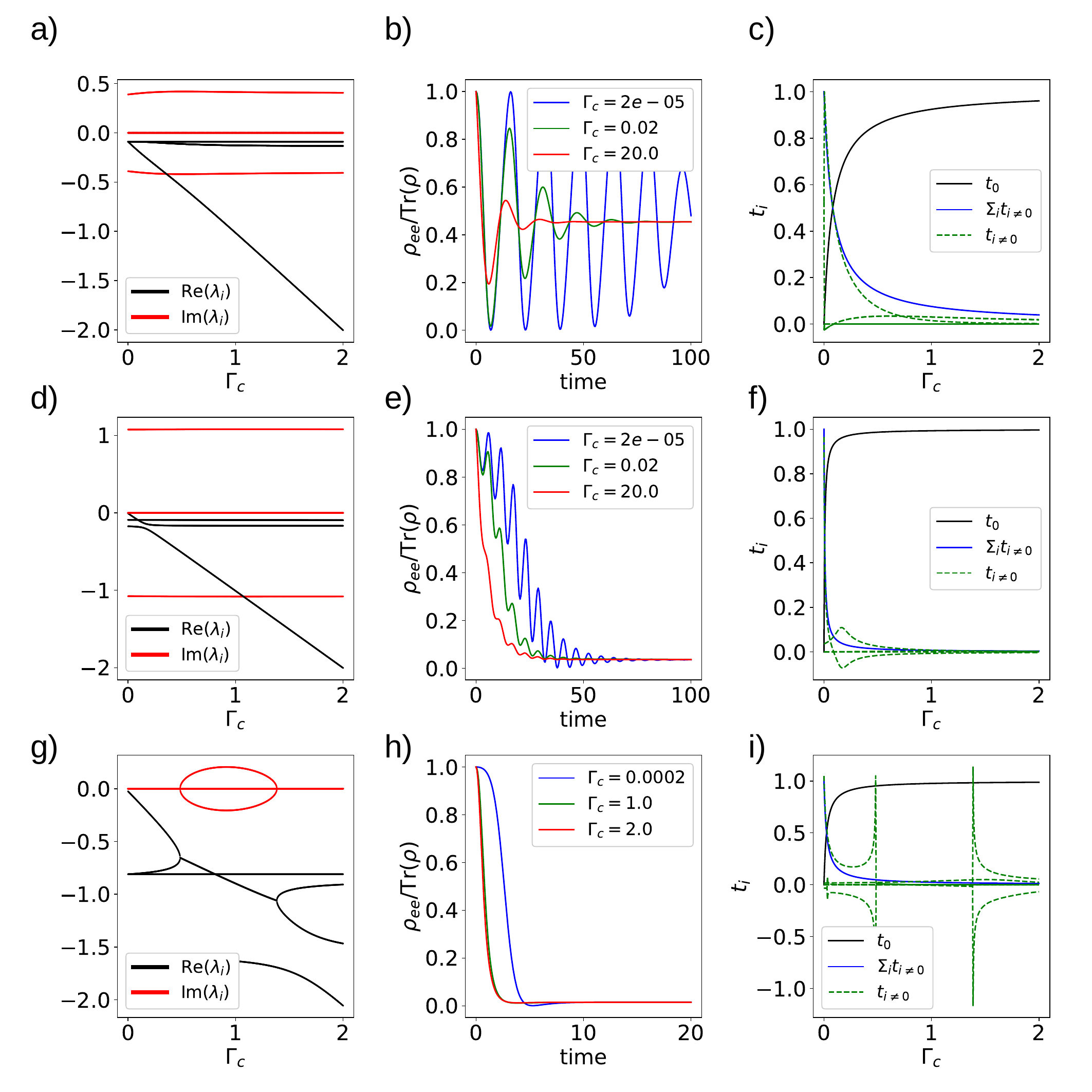}
    \caption{Left column: real (black) and imaginary (red) part of the eigenvalues as a function of $\Gamma_c$ for $V_{eg}=0.2$ and $\sqrt{\gamma/2\pi}=0.3$, on-resonance $\delta_e=0.001$ (top row) off-resonance $\delta_e=1.0$ (middle row) and a third set of parameters  $\delta_e=0.001$, $V_{eg}=0.1$ and $V_{ek}=0.9$ (bottom row). Middle column: evolution of the excited state population for a normalized density matrix taken at different values of $\Gamma_c$, for the previously mentioned set of parameters. Right column: Trace of the projectors $t_0$ and $\sum_i t_{i \neq 0}$ for the previously mentioned set of parameters (see main text)}
    \label{fig:eigenvalues}
\end{figure}

\end{widetext}


\subsection{M-level system}

The generality of the approach allows us to treat systems of arbitrary size, coupling connectivity, and in contact with a bath at finite temperature. We consider a second example for an $M$ level system, which is important in that it is able to support coherent population trapping points for certain values of the parameters (\cite{Finkelstein-Shapiro2019a}).  These are special conditions under which the population is trapped in the ground state coherently. It consists of two excited states connected to three ground states by radiative transitions in an $M$-like pattern. 
We allow relaxation from the excited to the ground states as well as incoherent excitation from ground to excited states to simulate a finite temperature bath. The Hamiltonian is:

\begin{equation}
\begin{split}
    H &= \delta_1 \ketbra{g_1}{g_1}+\delta_2 \ketbra{g_2}{g_2} + \delta_3 \ketbra{g_3}{g_3}\\
    & + (V_1^1 \ketbra{g_1}{e_1} + V_2^1 \ketbra{g_2}{e_1}\\
    & + V_2^2 \ketbra{g_2}{e_2} + V_3^2 \ketbra{g_3}{e_2}+ h.c.)\\
\end{split}
\end{equation}
where the $\delta_i$ are detunings and $V_i^j$ Hamiltonian couplings between the ground state $g_i$ and the excited state $e_j$. The jump operators for the different dissipative channels are $F_1=\sqrt{\gamma_{11}}\ketbra{g_1}{e_1}$, $F_2=\sqrt{\gamma_{12}}\ketbra{g_2}{e_1}$, $F_3=\sqrt{\gamma_{22}}\ketbra{g_2}{e_2}$, $F_4=\sqrt{\gamma_{23}}\ketbra{g_3}{e_2}$, $F_1'=\sqrt{\gamma'_{11}}\ketbra{e_1}{g_1}$, $F_2'=\sqrt{\gamma'_{21}}\ketbra{e_1}{g_2}$, $F_3'=\sqrt{\gamma'_{22}}\ketbra{e_2}{g_2}$, $F_4'=\sqrt{\gamma'_{32}}\ketbra{e_2}{g_3}$. The primed operators denote incoherent pumping whereas the non-primed operators are dissipative decay.

\begin{widetext}
\begin{figure}
    \centering
    \includegraphics[width=0.7\textwidth]{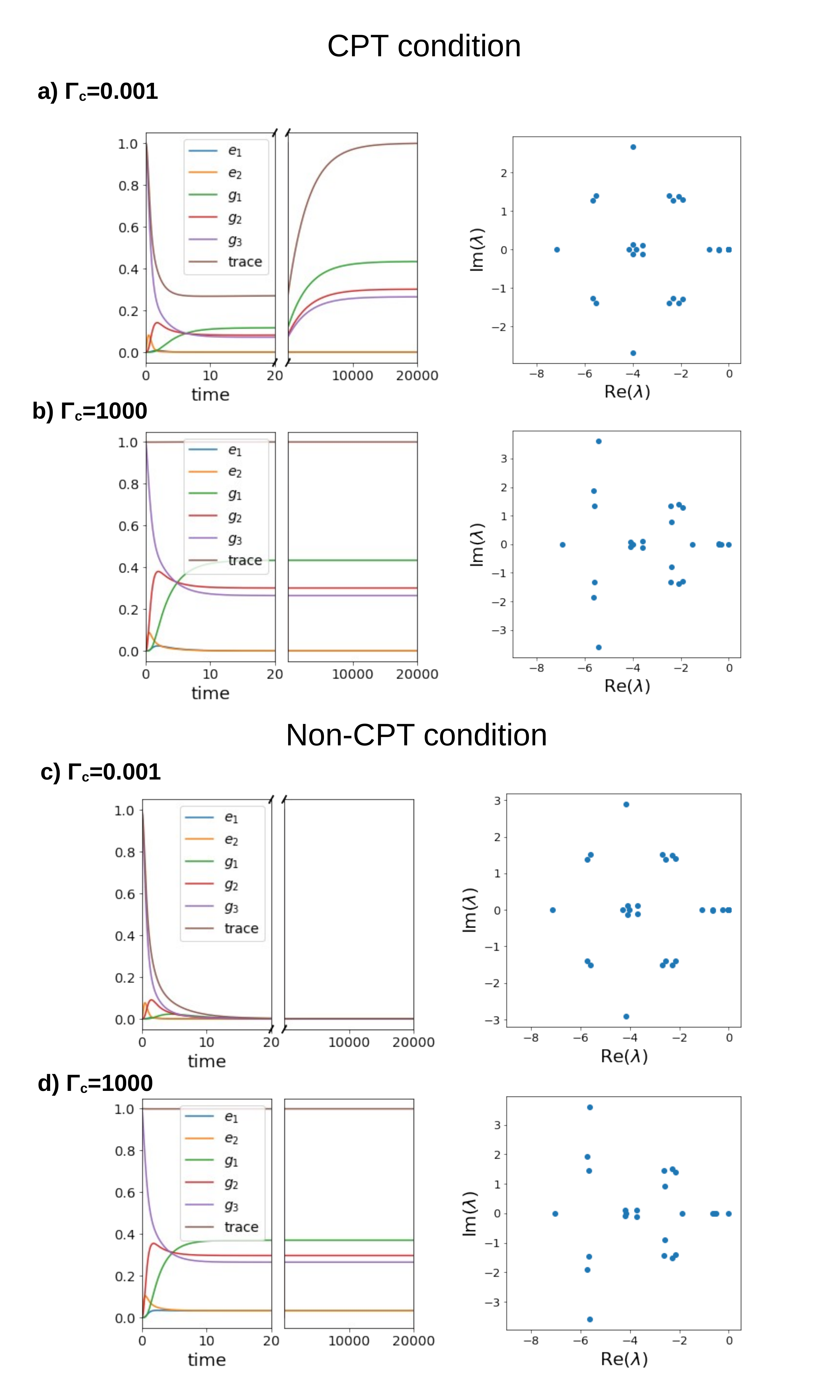}
    \caption{M-level system of two excited and three ground state. CPT condition corresponding to all detunings $\delta_i=-0.1$ and $V=$$V_{1}^{1}=1.0$
,$V_{2}^{1}=1.2$
,$V_{3}^{1}=0.0$
,$V_{1}^{2}=0.0$
,$V_{2}^{2}=1.5$
,$V_{3}^{2}=1.6$, 
and $\gamma_{ij}=2.0$, $\gamma'_{ij}=0.0$ for $\Gamma_c=0.001$ (a) and $\Gamma_c=1000$ (b). Non-CPT condition is obtained for the same parameters and with incoherent pumping corresponding to $\gamma'_{ij}=0.1$, for $\Gamma_c=0.001$ (c) and $\Gamma_c=1000$ (d).
}
    \label{fig:system_32}
\end{figure}
\end{widetext}

The coherent population trapping point occurs when all detunings are equal, and when no incoherent coupling is present \cite{Finkelstein-Shapiro2019a}. We first need to calculate the Lindblad operator and the Liouvillian constructed from the non-Hermitian Hamiltonian.
We construct the matrix $\tilde{M}$ according to Eq. \eqref{eq:M_matrix} to build the evolution operator. For the case of the two-level system, we made an effort to ascertain the number of poles which were expected to equal $-\Gamma_c$ and were thus singularities removed by the numerator $(z+\Gamma_c)$. This is unnecessary and we can construct the evolution operator without a detailed analysis of the poles. We plot in Figure \ref{fig:system_32} the ground and excited states for an initial state $\rho_S(0)=\ketbra{g_3}{g_3}$, as well as the trace of the density matrix, for several cases. First, we consider the parameters for coherent population trapping, obtained for $\delta_i=-0.1$, $V_{1}^{1}=1.0$
,$V_{2}^{1}=1.2$
,$V_{3}^{1}=0.0$
,$V_{1}^{2}=0.0$
,$V_{2}^{2}=1.5$
,$V_{3}^{2}=1.6$, $\gamma_{ij}=2.0$, $\gamma'_{ij}=0.0$ (zero temperature bath with no incoherent pumping). This set of parameters implies that $\text{ker}(L_\text{NH})=\text{ker}(L_\text{Lindblad})$ and they will only differ in the total trace since in the case of the non-Hermitian Hamiltonian some trace will be lost while the system reaches steady-state. We can see this clearly in Figures \ref{fig:system_32}.a and b which show the evolution and eigenvalues of $M$ for the coherent population trapping point for $\Gamma_c=0.001$ and $\Gamma_c=1000$. For small values of $\Gamma_c$, the system reaches a quasi-stationary state ($t<20$) where some of the trace remains in the continuum. However, because there is a CPT, on the timescale of $1/\Gamma_c$ the lost trace will leak back into the CPT steady-state until all of the particle density is back in the ground state. Only in the limit of $\Gamma_c \to 0$ does the trace remain in the continuum. Outside of the CPT, the trace remains in the continuum even when $\Gamma_c$ is finite. 

This behavior is in contrast with the case of a finite temperature bath ($\gamma'_{ij}=0.2$) where we have incoherent pumping and thus break the CPT condition. In this case the kernel of $\Lnh$ is empty and all of the particle density leaks out. In Figure \ref{fig:system_32}.c we observe that the trace vanishes ($t<20$) and is not recovered even at late times ($t>1/\Gamma_c$). All of the particle density escapes into the continuum sink and does not return. For Lindbladian dynamics modeled for $\Gamma_c=1000$ and shown in Figure  \ref{fig:system_32}.d, the trace is always conserved and now the excited states can become populated.

This example illustrates the role that the continuum plays in the case where the steady-state of $\Lnh$ and of $L_\text{Lindblad}$ are identical up to a normalization constant. The parameter $\Gamma_c$ sets a recurrence time for the lost density to return to the discrete partition, while outside of CPT the parameter $\Gamma_c$ controls the amount of trace that remains in the continuum. 


\clearpage

\section{Discussion}

We have strived to construct the simplest physical model to interpolate between non-Hermitian Hamiltonian and Lindbladian evolution in such a way that the steady-state trace varies continuously between 1 and 0. Projecting out an ancillary continuum of sublevels within a subsystem results in a non-Hermitian term that destroys the population and coherences of the discrete levels coupled to the continuum, and corresponds exactly to the non-Hermitian Hamiltonian term, and the second part of the Lindblad operator $\sum_i -\frac{1}{2} \{F_i^{\dagger} F_i,\rho\}_+$. It seems reasonable to restore the population lost in the continuum back to the ground state by a Lindblad operator. The simplicity of this model is seen in the compactness of $\Leff$ and the order of the nonlinearity in z, which is only quadratic. As a comparison, for a three-level system, removing one of the discrete levels from the explicit description adds a non-linearity in z of order five (the Liouville space of three levels is nine while that of two levels is four, so that we need a dependence on z to the fifth power to capture all the dynamical variables).

The structure of the model is inspired by work on Hamiltonians with continuous manifolds from atomic physics (i.e. Fano Hamiltonians \cite{Fano1961}) adapted to dissipative environments \cite{Finkelstein2015,Finkelstein2018,Finkelstein2015,Finkelstein2016-1,Finkelstein-Shapiro2020}. 
It can be thought of as a simple model for systems consisting of an extended structure (semiconductor or metal) with molecular adsorbates or defects \cite{Finkelstein_plexciton_theory_arxiv}, and also as a first-order approximation to the adiabatic elimination of discrete excited states in the limit of very fast dissipation back to the ground state \cite{Finkelstein-Shapiro2020}.
It can be used to describe real systems such as metals coupled to semiconductors \cite{Zhang2006,Zhang2011}, molecules injecting charge to semiconductors (\cite{Wang2005, Petersson2000}, and waveguides \cite{Baernthaler2010}. Models of Fano with dissipation were of interest from the early days of photoionization \cite{Agarwal1982,Rzazewski1983} and have remained important up to the latest experiments in attosecond spectroscopy \cite{Busto2022}. 
Ultrafast experiments on semiconductors with Fano structures have investigated the dephasing mechanisms \cite{Siegner1995b,Glutsch19951} and it is expected that newer techniques will help to further elucidate the complex relaxation dynamics and lineshapes \cite{FinkelsteinShapiro2018}.

We have discussed for the M-level system the effect of having coherent population trapping points. There are other conditions of interest in open quantum systems. Exceptional points - the coalescence of eigenvalues and eigenvectors - are in general different for non-Hermitian or Lindblad operators \cite{Minganti2019}. 
The non-monotonicity of the eigenvalues as a function of $\Gamma_c$ is further proof that the structure of the eigenvalues and hence exceptional points of both non-Hermitian and Lindblad limits are different, and that some non-trivial behavior could be found between the two. 

The proposed interpolation between Lindblad and non-Hermitian evolution works for all times, all coupling strengths and finite temperatures. It will apply for all cases when one can write a valid Lindblad operator, including those with pure dephasing where a non-Hermitian equivalent is not obvious. 
We have assumed throughout that the extended matrix $\tilde{M}$ has semisimple eigenvalues, and in particular that the geometric and algebraic multiplicities of $\lambda = -\Gamma_c$ are the same. This is true for the two-level and $M$-level system studied, however we have not proved it to be the case in general. We will approach this point in a future work.

Finally, we note that we begin with two maps which are Markovian, however we end up with a non-Markovian map. The expressions developed here assume that $\sQ \rho(0) = 0$. Since in general at any given time $t>0$ this is no longer true, we have that $U(t+\tau) \neq U(t)U(\tau)$. \newline 

\section{Conclusion}

We have proposed a more general generator of dynamics that allows for a continuous transformation between pure decay dynamics obtained from a non-Hermitian Hamiltonian to the trace-preserving dynamics induced by Lindblad operators. This effective operator is rooted in a microscopic derivation using an ancillary continuum, is energy dependent and so the inverse Laplace transform of its resolvent is not trivial. 
As long as the decay rates from all the ancillary continua back to the system are the same, the nonlinearities are quadratic and a procedure is proposed to obtain the exact evolution operator. To this end we resort to a copy of Liouville space whose full usefulness and meaning remains to be explored. Both Non-Hermitian and Lindblad maps are extensively used in open quantum systems from phase transitions to spectroscopical observables. Our result presents a fundamental connection between them, and also opens a new avenue in the analysis of maps that describe systems exchanging energy and particle density with their surroundings. 

\section{Acknowledgements}

D.F.S thanks Prof. Arne Keller for useful discussions and acknowledges PAPIIT grant IA202821. 

\bibliography{Fano,dark-states,adiabElimin,adiabElimBipartNotes,NH_L}

\appendix

\setcounter{section}{0} 
\renewcommand{\thesection}{A} 

\setcounter{figure}{0} 
\renewcommand{\thefigure}{\thesection.\arabic{figure}}

\numberwithin{equation}{section} 
\section{Appendix A. Derivation of the operators for the mixed dynamics.}

We derive a solution for the effective Liouvillian in the mixed model of Figure \ref{fig:microscopic_models}.c. We solve the more general case where incoherent pumping is also possible, so that we use two continua, one associated to the decay channel and one with the pumping channel. We can revert to a simple decay by putting the couplings for the incoherent pumping channel to zero. 

\begin{figure}[ht]
\includegraphics[width=0.5\textwidth]{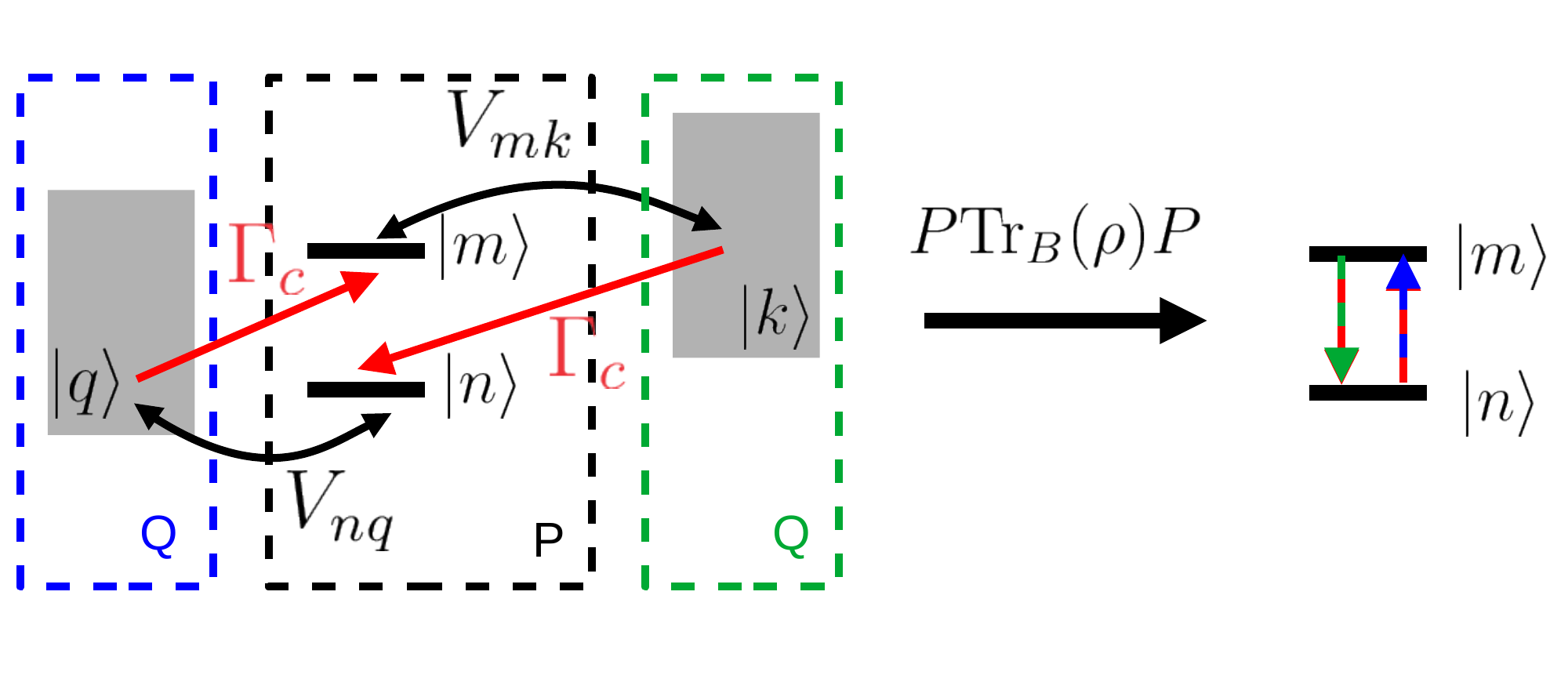}
\caption{\label{fig:FanoDissip} Energy levels and transitions between two levels with incoherent transitions mediated by continua. A dissipative pathway from $m$ to $n$ with rate $\gamma_{mn}$ is replaced by a Hamiltonian transition from $m$ to a continuum $k$ with coupling strength $V_{mk}=\sqrt{\gamma_{mn}/2\pi}$ and a dissipative rate from the continuum $k$ back to $n$ with rate $\Gamma_c$. The incoherent pumping from $n$ to $m$ with rate $\gamma_{nm}$ is replaced by a Hamiltonian transition from $n$ to a continuum $q$ with coupling strength $V_{nq}=\sqrt{\gamma_{nm}/2\pi}$ and a dissipative rate from the continuum $q$ back to $m$ with rate $\Gamma_c$. }
\label{fig:finite_bath}
\end{figure}

We want to describe a pair of levels $n$ and $m$ with possible dissipative transitions between the two. 
The recipe used is to intersperse a continuum $\{\ket{k}\}$ between $m$ and $n$ to mediate a dissipative transition in Lindblad form from $m$ to $n$ and another continuum $\{\ket{q}\}$ to mediate a dissipative transition in Lindblad form from $n$ to $m$. We focus on pairs of levels without loss of generality as all transitions are pairwise. We do not explicitly include Hamiltonian couplings between $m$ and $n$ as these can be added later without affecting the result (Figure \ref{fig:finite_bath}). 

The Hamiltonian is
\begin{equation}
\begin{split}
    H &= H_{\text{discrete}}+\sum _i H_{\text{continuum,i}}\\
    &+\sum _i H_{\text{discrete-continuum,i}} \\
    H_\text{discrete} &=  \omega_n\ketbra{n}{n}+\omega_m\ketbra{m}{m} \\
    H_\text{continuum,1} &= \int dk \omega_k \ketbra{k}{k} \\
    H_\text{continuum,2} &= \int dq \omega_q \ketbra{q}{q} \\
    H_\text{discrete-continuum,1} &= \int dk (V_{mk}\ketbra{m}{k} + h.c.) \\
    H_\text{discrete-continuum,2} &= \int dq (V_{nq}\ketbra{n}{q} + h.c.)
\end{split}
\end{equation}
As before we can build the conservative Liouvillian in Liouville space $L_H = -i[1\otimes H - H^* \otimes 1]$ and add the dissipative transitions corresponding to the jump operators $F_{mn}^{(k)} = \sqrt{\Gamma_c} \ketbra{n}{k}$ and $F_{nm}^{(q)} = \sqrt{\Gamma_c} \ketbra{m}{q}$, where we have already chosen the same relaxation rate $\Gamma_{c}$ for both ancillary continua back to the discrete manifold.

We need to project out the continuum in partition $\sQ$. Since we only need to calculate $\sP U(t) \sP$, this is equivalent to calculating $\sP G(z) \sP$. We can write a Lippman-Schwinger expansion $G = G_0 + G_0 W G$ if we re-express the Lindblad operator as
\begin{equation}
    L = L_0+W
\end{equation}
where $L_0 = \sP L \sP + \sQ L \sQ$ and $W = \sP L \sQ + \sQ L \sP$, $G_0(z) = (z-L_0)^{-1}$ and $G(z) = (z-L)^{-1}$. Using $\sP+\sQ = 1$ we can write $G(z) = G_0(z) + G_0(z)WG(z) = G_0(z) + G_0(z)(\sP+\sQ)W(\sP+\sQ)G(z)$, and projecting onto the different subspaces we obtain a set of four relations
\begin{equation}
    \begin{split}
        \sP G(z) \sP = \sP G_0(z) \sP + \sP G_0(z) \sP W \sQ G(z) \sP \\
        \sP G(z) \sQ = \sP G_0(z) \sP W \sQ G(z) \sQ \\
        \sQ G(z) \sP = \sQ G_0(z) \sQ W \sP G(z) \sP \\
        \sQ G(z) \sQ = \sQ G_0(z) \sQ + \sQ G_0(z) \sQ W \sP G(z) \sQ \\
    \end{split}
\end{equation}
From which we recover after some algebra
\begin{equation}
    \sP G(z) \sP = \sP G_0(z) \sP + [\sP G_0(z) \sP] [ \sP W \sQ G_0(z) \sQ W \sP  ] [\sP \sP G(z) \sP]
    \label{eq:LS_in_P}
\end{equation}
which we have written Eq. \eqref{eq:LS_in_P} suggestively to identify it with a Lippman-Schwinger expansion contained in $\sP$ in order to identify an effective operator (Eq. \eqref{eq:Leff}) such that $\sP G(z) \sP = [z-\Leff]^{-1}$. 
In the particular case of two levels and two continua, 
 the needed Feshbach projectors are $P = \ketbra{n}{n} + \ketbra{m}{m}$, $Q = \int dk \ketbra{k}{k}+\int dq \ketbra{q}{q}$ and $\sP = P \otimes P$, $\sQ = 1-\sP$, we can calculate $\sP L \sQ G_0(z) \sQ L \sP$. The explicit form of the operators needed are
\begin{equation}
\begin{split}
\sQ G_0(z) \sQ &= \int dk \int dk' \frac{\ketbra{kk'}{kk'}}{z-i\omega_{kk'}+\Gamma_c} \\
&+ \int dq \int dq' \frac{\ketbra{qq'}{qq'}}{z-i\omega_{qq'}+\Gamma_c}\\
&+ \int dk \frac{\ketbra{km}{km}}{z-i\omega_{km}+\Gamma_c/2} + h.c. \\
&+ \int dq \frac{\ketbra{qn}{qn}}{z-i\omega_{qn}+\Gamma_c/2} + h.c. \\
\end{split}
\end{equation}
and
\begin{equation}
\begin{split}
    \sP L \sQ &= \Gamma_c\ketbra{nn}{kk} \\&+ \Gamma_c\ketbra{mm}{qq} \\ &+\sum_{a=m,n}\left[iV_{mk}\ketbra{am}{ak}+h.c.\right] \\ &+\sum_{a=m,n}\left[iV_{nq}\ketbra{an}{aq}+h.c.\right] \\
    \sQ L \sP &=  \sum_{a=m,n}\left[iV_{mk}\ketbra{ak}{am}+h.c.\right] \\ &+\sum_{a=m,n}\left[iV_{nq}\ketbra{aq}{an}+h.c.\right] \\
    \end{split}
\end{equation}
 Considering a wideband approximation where $\omega_k = k$, $\omega_q = q$, the integrals in $\sP L \sQ G_0(z) \sQ L \sP$ are readily evaluated to $\pi$ and we can obtain  (see also \cite{FinkelsteinShapiro2018,Finkelstein-Shapiro2020}).:
\begin{equation}
\begin{split}
    &\Leff(z)-\sP L \sP+ =  \sP L \sQ G_0(z) \sQ L \sP \\
    &= 2\pi \abs{V_{mk}}^2 \left(\frac{\Gamma_c}{z+\Gamma_c} \ketbra{n}{m} \otimes \ketbra{n}{m} \right. \\& \left. -  \frac{1}{2}\left( 1\otimes \ketbra{m}{m} + \ketbra{m}{m} \otimes 1 \right)\right) \\
    &+ 2\pi \abs{V_{nq}}^2 \left(\frac{\Gamma_c}{z+\Gamma_c} \ketbra{m}{n} \otimes \ketbra{m}{n} \right. \\& \left. -  \frac{1}{2}\left( 1\otimes \ketbra{n}{n} + \ketbra{n}{n} \otimes 1 \right) \right)
\end{split}
\end{equation}
We set the values for the dissipation rate going from $m$ to $n$ as $\gamma_{mn} = 2 \pi \abs{V_{mk}}^2$ and from $n$ to $m$ as $\gamma_{nm} = 2 \pi \abs{V_{nq}}^2$. To get the expression into the final form of Eq. \eqref{eq:Leff}, we relabel each transition between pairs of states by the global label $i$ and re-express the generalized operator in Hilbert space.

\setcounter{section}{0} 
\renewcommand{\thesection}{B} 

\section{Appendix B. Structure of the determinant}

We find that under some conditions the extended matrix $\tilde{M}$ has poles at $\lambda= -\Gamma_c$, and these simple poles are cancelled by the numerator so that the number of poles of the generalized resolvent is less than the number of distinct eigenvalues of $\tilde{M}$. As mentioned, it is not necessary to know how many poles are canceled as the projection operator will vanish when $\lambda= -\Gamma_c$. 

We can nonetheless make explicit the number of poles that are equal to $-\Gamma_c$ by looking more in detail to a two-level system. \newline

\textbf{Two-level system with pure decay.} The aim is to find the zeros of the determinant of $z-\tilde{M}$:
\begin{equation}
\begin{split}
    &\text{det}\left(z-\tilde{M}\right) \\ &= \text{det}\left( \begin{bmatrix}
        z & -1 \\
        -\Gamma_c \LD & z-\LD+\Gamma_c + J
        \end{bmatrix} \right)
        \end{split}
\end{equation}
where we have implied $z=\openone_8z$ where $\openone_8$ is the identity matrix in eight dimensions. In what follows, where a constant appears we assume it is proportional to the identity in the relevant dimension. The extended matrix can be written pictorially as:
\begin{equation}
\begin{split}
    z-\tilde{M} &= \begin{bmatrix} 
        \begin{array}{c c c c | c c c c}
        \boxG & 0 & 0 & 0 & \boxG & 0 & 0 & 0 \\
        0 & \boxG & 0 & 0 & 0 & \boxG & 0 & 0 \\
        0 & 0 & \boxG & 0 & 0 & 0 & \boxG & 0 \\
        0 & 0 & 0 & \boxG & 0 & 0 & 0 & \boxG  \\ \hline
        0 & \boxG & \boxG & \boxB & \boxG & \boxG & \boxG & 0 \\
        \boxG & \boxG & 0 & \boxG & \boxG & \boxG & 0 & \boxG \\
        \boxG & 0 & \boxG & \boxG & \boxG & 0 & \boxG & \boxG \\
        0 & \boxG & \boxG & \boxG & 0 & \boxG & \boxG & \boxG  \\
    \end{array}
    \end{bmatrix}
    \end{split}
    \label{eq:boxes_1}
\end{equation}
where we have explicitly separated blocks of the size of the original Liouville space. We denote by $\boxB$ the element of $J$. 

We begin by looking at the determinant of the simpler matrix containing elements \boxG \quad only. 
\begin{equation}
\begin{split}
    D_0 &= \det(z-M') \\&= \det \left(  \begin{bmatrix}
           z & 1 \\
           -\Gamma_c (\LD -J) & z-(\LD-J - \Gamma_c)
    \end{bmatrix} \right)
    \end{split}
\end{equation}
where $M'$ is a matrix similar to $M$ but Liouvillians built from non-Hermitian Hamiltonians only. 
We can express simply the determinant using the identity $\det \left(\begin{bmatrix}
    A & B \\ C & D
\end{bmatrix} \right) = \det(AD-BC)$ where we have used the fact that $C$ and $D$ commute. Then $\det(M') = \det( z(z-(\LD-J) + \Gamma_c) - \Gamma_c(\LD-J)   ) = \det( (z+\Gamma_c)(z-(\LD-J)) ) = (z+\Gamma_c)^4\det(z-(\LD-J))$.
The determinant of the full matrix $\tilde{M}$ can be expressed as the contribution $D_0$ and the additional permutations that include the term arising from the operator $J$ marked as \boxB. 
\begin{equation}
\begin{split}
\det(\tilde{M}) &= (z+\Gamma_c)^4\det(z-(\LD-J)) + \text{Permutations}(\boxB) \\
 &= (z+\Gamma_c)^4\det(z-(\LD-J)) + (z+\Gamma_c)^3\times...  \\
\end{split}
\end{equation}
Since the permutations involving $J$ replace one order of $z$ from the upper left block matrix, and the permutations of the remaining three powers of $z$ with the \boxG \quad elements shift them by $+\Gamma_c$, this exta contribution is proportional to $(z+\Gamma_c)^3$. \newline

\begin{figure}[h]
    \centering
    \includegraphics[width=0.5\textwidth]{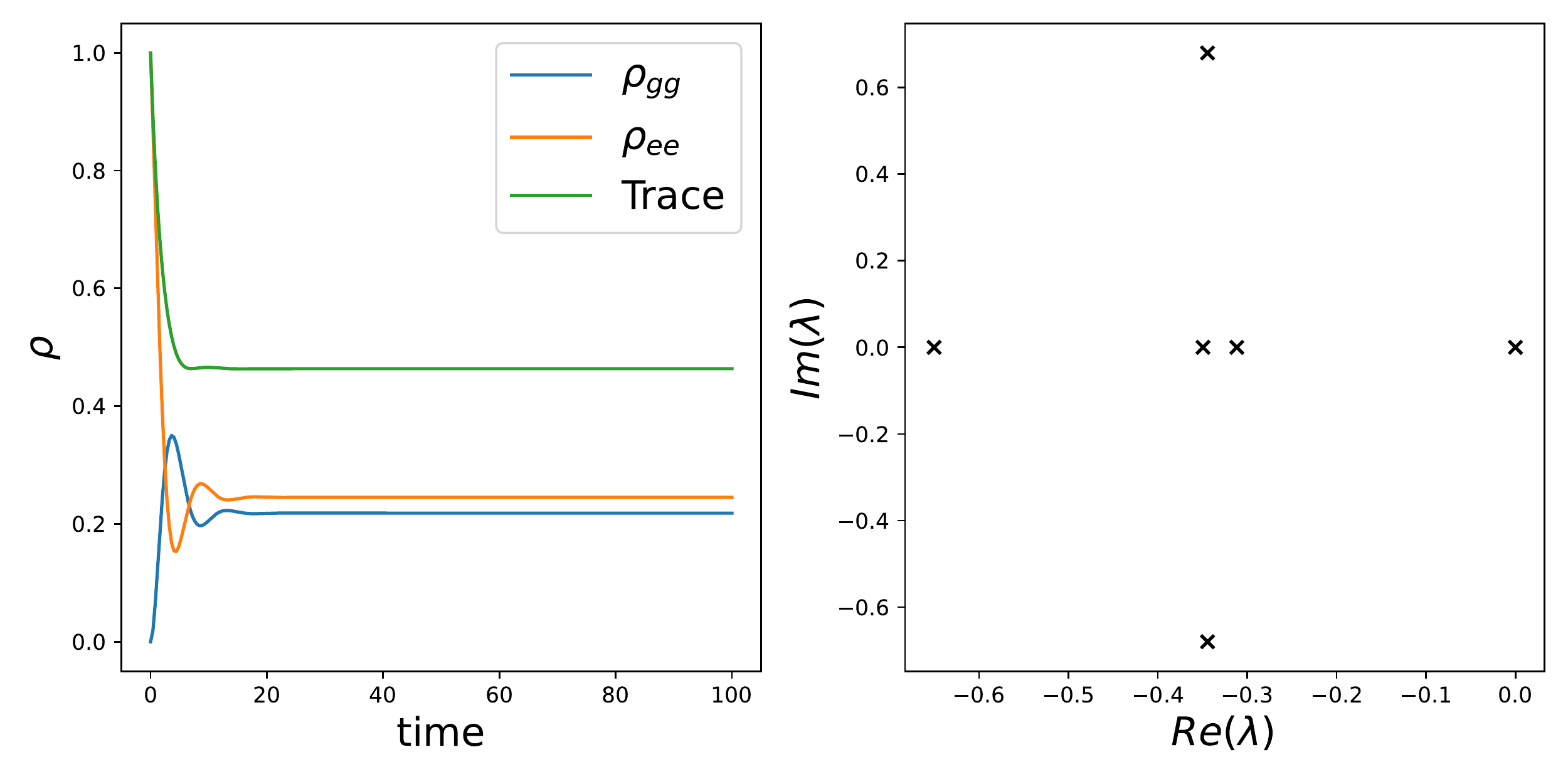}
    \caption{Evolution of $\rho_{gg}$ and $\rho_{ee}$ (left) and the pole structure of the generalized resolvent (right) for a two-level system coupled to a bath at finite temperature. Chosen parameters are $\delta_e = 0.01$, $V_{eg}=0.3$, $\gamma_{eg}=0.15$, $\gamma_{ge}=0.2$ and $\Gamma_c=0.3$}
    \label{fig:evolution_finite_T_bath}
\end{figure}

\begin{figure}[h]
    \centering
    \includegraphics[width=0.5\textwidth]{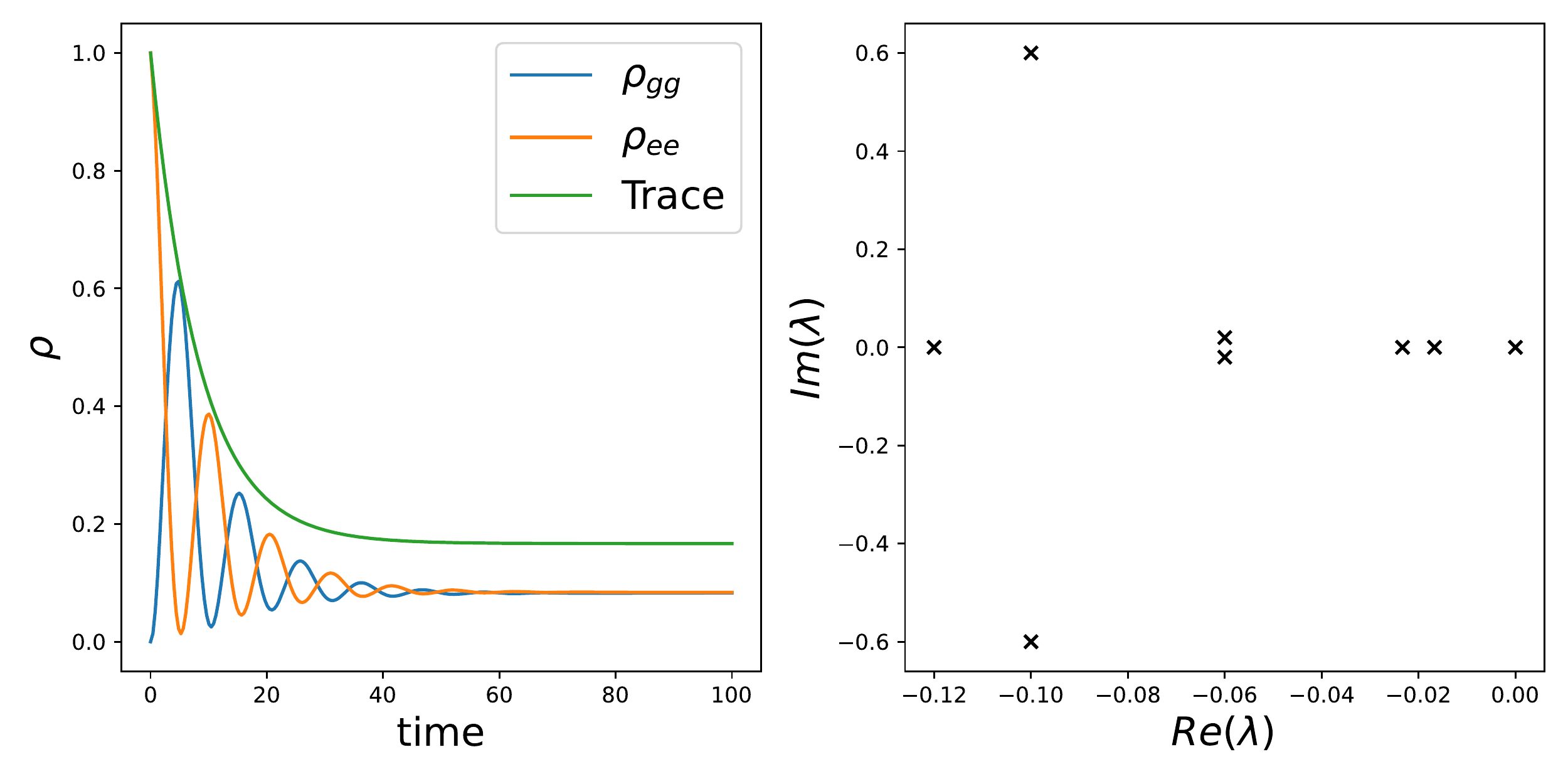}
    \caption{Evolution of $\rho_{gg}$ and $\rho_{ee}$ (left) and the pole structure of the generalized resolvent (right) for a two-level system with pure dephasing. Chosen parameters are $\delta_e = 0.01$, $V_{eg}=0.3$, $\gamma_{z}=0.1$, and $\Gamma_c=0.02$}
    \label{fig:evolution_pure_dephasing}
\end{figure}

\textbf{Two-level system with additional dissipative transitions.} In the case of a bath at finite temperature, we have to calculate the determinant including also incoherent pumping marked as \boxS
\begin{equation}
 \begin{bmatrix} 
        \begin{array}{c c c c | c c c c}
        \boxG & 0 & 0 & 0 & \boxG & 0 & 0 & 0 \\
        0 & \boxG & 0 & 0 & 0 & \boxG & 0 & 0 \\
        0 & 0 & \boxG & 0 & 0 & 0 & \boxG & 0 \\
        0 & 0 & 0 & \boxG & 0 & 0 & 0 & \boxG  \\ \hline
        \boxG & \boxG & \boxG & \boxB & \boxG & \boxG & \boxG & 0 \\
        \boxG & \boxG & 0 & \boxG & \boxG & \boxG & 0 & \boxG \\
        \boxG & 0 & \boxG & \boxG & \boxG & 0 & \boxG & \boxG \\
        \boxS & \boxG & \boxG & \boxG & 0 & \boxG & \boxG & \boxG  \\
    \end{array}
    \end{bmatrix}
    \label{eq:boxes_1}
\end{equation}
and the permutations involving both \boxS \quad and \boxB \quad are now proportional to $(z+\Gamma_c)^2$ so that the resolvent has six poles. The dynamics and pole structure of a two level system with a decay and incoherent pumping is shown in Figure \ref{fig:evolution_finite_T_bath} where we can appreciate both the loss of trace as well as the existence of six poles.
The case of pure dephasing (Fig. \ref{fig:evolution_pure_dephasing}) involves all eight poles. 

\end{document}

\section{For figures}

\begin{equation*}
    \mathcal{H}_S
\end{equation*}

\begin{equation*}
    \mathcal{H}_A
\end{equation*}

\begin{equation*}
    \mathcal{H}_B
\end{equation*}

\begin{equation*}
    \oplus
\end{equation*}

\begin{equation*}
    \otimes
\end{equation*}

\end{document}